\documentclass{jps-cp}
\usepackage{txfonts,microtype} 

\title{Altermagnetism in MnF$_2$: Band Splitting and Its Physical Consequences}

\author{Igor \textsc{Solovyev}$^{1}$}

\inst{$^{1}$Research Center for Materials Nanoarchitectonics (MANA), National Institute for Materials Science (NIMS), 1-1 Namiki, Tsukuba, Ibaraki 305-0044, Japan}

\email{SOLOVYEV.Igor@nims.go.jp}

\recdate{June 18, 2026}

\abst{MnF$_2$ is widely regarded as a candidate altermagnet, but the magnitude and implications of its altermagnetic band splitting remain debated. Using electronic-structure calculations, we construct minimal models that capture the magnetic and electronic properties of MnF$_2$. These models show that the parameters governing the chiral magnon splitting and the spin splitting of the electronic bands are relatively small. Moreover, the electronic system lies in the strong-coupling regime, where most magnetic properties are controlled by the ratio $t/U$ between the characteristic hopping amplitude $t$ and the large on-site Coulomb repulsion $U$. Consequently, all exchange interactions scale as $1/U$, so a small altermagnetic hopping $\delta t$ produces only a proportionally small exchange term. Upon doping, the altermagnetic contribution to the anomalous Hall effect is likewise suppressed, being smaller than the conventional (non-altermagnetic) contribution by a factor of order $\delta t/U$. In contrast, the behavior of the conductivity tensor $\hat{\sigma}(\omega)$ at $\hbar \omega \sim U$ differs qualitatively, because $\delta t$ enters the energies of interband optical transitions \emph{directly} rather than through the reduced ratio $\delta t/U$. This contribution strongly reshapes $\hat{\sigma}(\omega)$, leading to a dramatic enhancement of the magneto-optical response.}

\kword{altermagnetic band splitting, exchange interactions, anomalous Hall effect, magneto-optical effect, electronic structure calculations, minimal models}

\begin{document}
\maketitle

\section{\label{sec:intro} Introduction}
\par There has been growing interest in unconventional antiferromagnetism realized in materials where the antiferromagnetic (AFM) and chemical unit cells coincide~\cite{Dzyaloshinskii1991}. Such AFM order can break time-reversal symmetry ($\mathcal{T}$) globally, giving rise to weak ferromagnetism~\cite{Dzyaloshinskii_weakF}, piezomagnetism~\cite{DzyaloshinskiiPM}, and magnetoelectricity~\cite{DzyaloshinskiiME}, phenomena introduced by Dzyaloshinskii in the late 1950s within the framework of phenomenological Landau theory. These ideas were further developed by Turov in a series of publications~\cite{TurovShavrov,TurovUFN} and summarized in his monograph~\cite{TurovBook}. On the classification side, Turov proposed dividing unconventional antiferromagnets into centrosymmetric and centroantisymmetric classes, depending on whether spatial inversion $\mathcal{I}$ enters the magnetic group directly or only in combination with $\mathcal{T}$. The former class includes weak ferromagnetism and piezomagnetism, whereas the latter is associated with magnetoelectricity. The presence of inversion symmetry plays a crucial role in stabilizing nearly collinear antiferromagnetism, since in its absence the Dzyaloshinskii-Moriya (DM) interaction typically drives the system toward long-wavelength spin-spiral or skyrmionic textures~\cite{Dzyaloshinskii1964,bog2}. In other words, we are dealing with antiferromagnetism on an \emph{antipolar} structure~\cite{HiddenFM}, where inversion symmetry is broken on individual bonds but the bonds are arranged in an antipolar fashion, preserving the global inversion center. From the viewpoint of physical properties, Turov also formulated the key symmetry principles governing a wide range of kinetic, optical, and acoustic responses controlled by macroscopic time-reversal symmetry in these unconventional antiferromagnets~\cite{TurovBook}.

\par With time, first-principles calculations based on density functional theory have come to play an essential role in elucidating the microscopic origin of these phenomena. In particular, the DM interactions responsible for weak ferromagnetism~\cite{Dzyaloshinskii_weakF,Moriya_weakF} have now become accessible to quantitative evaluation within such approaches~\cite{PRL96,PRB2014,SzilvaRevModPhys,review2024}. Early first-principles studies of the antisymmetric optical conductivity in centrosymmetric antiferromagnets further revealed that it correlates much more strongly with the net orbital magnetization than with weak spin ferromagnetism~\cite{PRB1997}. Furthermore, first-principles calculations also uncovered a spin splitting of antiferromagnetic bands~\cite{Sandratskii1981,Noda,Okugawa,HayamiJPSJ,Naka}, formally predating the modern era of altermagnetism~\cite{Naka_Spintronics}.

\par The term ``altermagnetism'' was introduced independently in 2022 to denote a putative Type-III magnetic phase, distinct from conventional ferromagnetism and antiferromagnetism. It is characterized by crystal-rotation symmetries that connect opposite-spin sublattices, producing alternating spin splitting of electronic bands in reciprocal space and robust time-reversal-symmetry breaking even in the absence of relativistic spin-orbit (SO) coupling~\cite{SmejkalPRX1,MazinPRX1}. However, this proposed Type-III phase is not fundamentally distinct from the centrosymmetric antiferromagnetism introduced by Turov decades ago. Indeed, most of the 221 altermagnetic candidates listed in the recent review~\cite{LingBai} are centrosymmetric antiferromagnets in Turov’s classification. What was genuinely new in the recent discussion is the explicit recognition of altermagnetic band splitting -- an effect outside the scope of the phenomenological Landau theory of Refs.~\cite{Dzyaloshinskii_weakF,DzyaloshinskiiPM,DzyaloshinskiiME,TurovShavrov,TurovUFN,TurovBook} -- rather than any fundamental novelty in the underlying magnetic symmetry. Therefore, altermagnetic band splitting should be viewed simply as one of the possible microscopic manifestations that centrosymmetric antiferromagnets can host. An important question, then, is how this band splitting intertwines with other physical properties: in particular, what role it plays in magnon excitations, the anomalous Hall effect (AHE), and the magneto-optical response.

\par MnF$_2$ has attracted considerable attention as a candidate altermagnet, but its chiral magnon modes exhibit nearly vanishing splitting~\cite{MoranoMnF2,FaureMnF2}, which has sparked a discussion about the importance of this splitting for the physical properties of MnF$_2$ and other centrosymmetric antiferromagnets. In this article, using first-principles electronic-structure calculations, we argue that altermagnetic band splitting makes only a minor contribution to the magnon spectrum and to the anomalous Hall effect (AHE) upon doping. However, it can strongly enhance the magneto-optical response. The rest of the article is organized as follows. In Sec.~\ref{sec:str}, we briefly discuss the crystal structure and basic electronic structure of MnF$_2$. In Sec.~\ref{sec:ex}, we analyze the magnetic interactions extracted from the electronic-structure calculations and the resulting magnon dispersion. In Sec.~\ref{sec:effel}, we turn to the effective electronic model constructed to reproduce these magnetic interactions and examine the behavior of the AHE, optical conductivity, and magneto-optical response.  In Sec.~\ref{sec:summary}, we summarize our findings.

\section{\label{sec:str} Crystal structure and basic electronic structure}
\par
\begin{figure}[tbh]
\centering
\includegraphics[width=15.0cm]{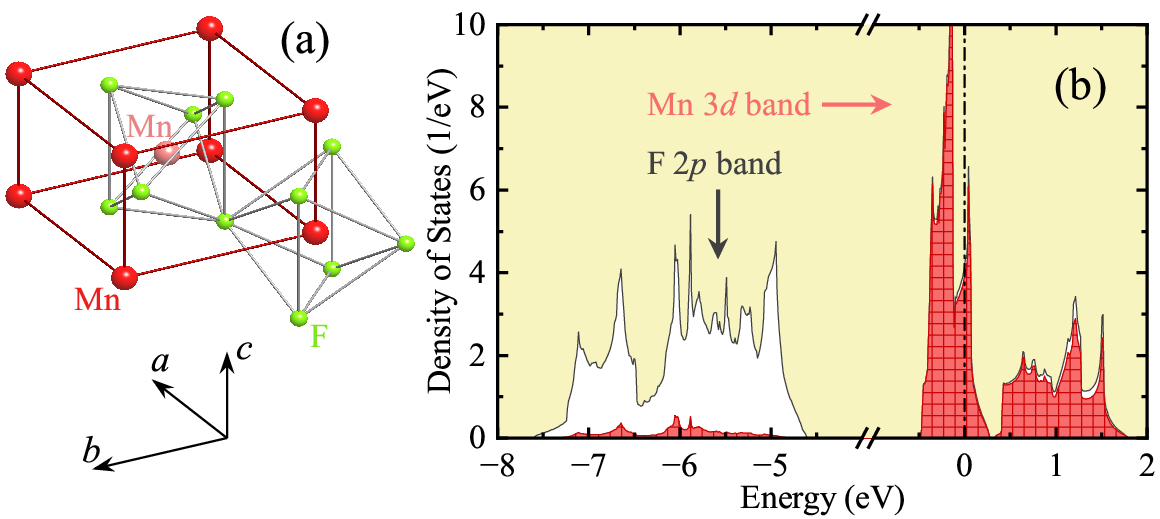}
\caption{(a) Fragment of the crystal structure of MnF$_2$, illustrating the relative arrangement of the MnF$_6$ octahedra. Mn atoms belonging to different magnetic sublattices are shown in different colors. (b) Total (white) and Mn $3d$ (red) densities of states calculated within the local-density approximation. The Fermi level is set to zero energy.}
\label{fig:str}
\end{figure}

 MnF$_2$ adopts the rutile structure (the space group $P4_{2}/mnm$, No. 136) with lattice parameters $a = 4.8736$~\AA~and $c=3.3$~\AA \, (see Fig.~\ref{fig:str}(a)). The F atom occupies the position $(0.305,0.305,0)$~\cite{MnF2}. Hereafter, all coordinates are given in units of $a$ and $c$. $P4_{2}/mnm$ comprises $\mathcal{I}$, the fourfold screw rotation $\{ \mathcal{C}_{4z} | \mathbf{t} \}$, and the glide-mirror operation $\{ m_{x} | \mathbf{t} \}$ with $\mathbf{t}=(\frac{1}{2},\frac{1}{2},\frac{1}{2})$.
\noindent
\noindent Thus, $\mathcal{I}$ maps each Mn sublattice onto itself, whereas $\{ \mathcal{C}_{4z} | \mathbf{t} \}$ and $\{ m_{x} | \mathbf{t} \}$ interchange the two sublattices. The Mn-F-Mn angle is $129^{\circ}$, which breaks inversion symmetry at the bond center and therefore allows finite DM interactions between the Mn sublattices. The screw-axis rotation $\{ \mathcal{C}_{4z} | \mathbf{t} \}$ allows altermagnetic band splitting~\cite{SmejkalPRX1}, but the symmetry alone does not determine the magnitude of the splitting.

\par The electronic structure in the local-density approximation (LDA) is extremely simple: the valence manifold consists of the F $2p$ and Mn $3d$ bands, which are separated by a large energy gap of about $4$ eV, as shown in Fig.~\ref{fig:str}(b).

\section{\label{sec:ex} Magnetic interactions}
\par
\begin{figure}[tbh]
\centering
\includegraphics[width=15.0cm]{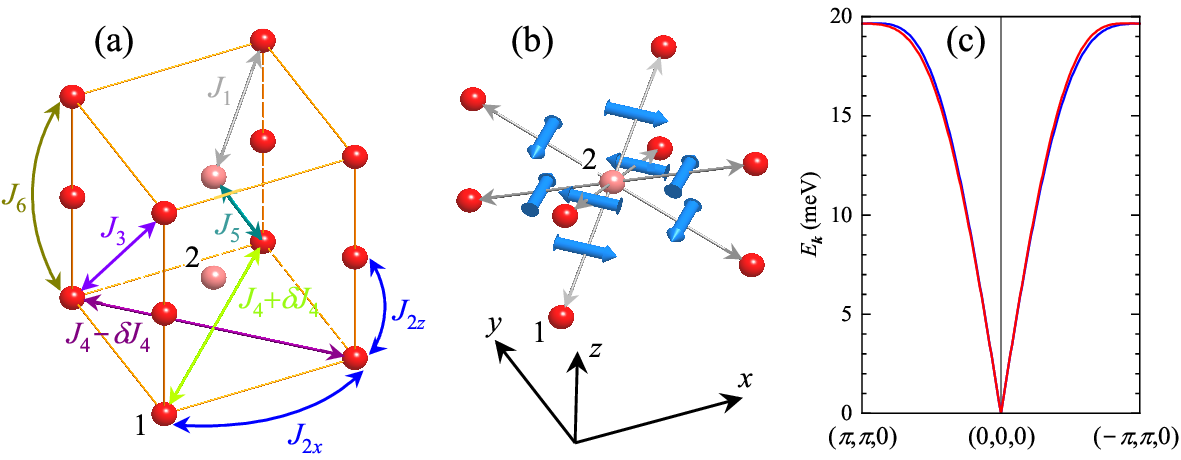}
\caption{(a) Main parameters of the isotropic exchange interactions. Mn atoms belonging to sublattices 1 and 2 are shown in different colors. The interactions on sublattice 2 have the same form, except that the bonds $J_{4} \pm \delta J_{4}$ must be interchanged. (b) Dzyaloshinskii-Moriya interaction vectors (blue arrows) acting between the sublattices. The corresponding bond directions are indicated in grey. (c) Example of the magnon dispersion for the N\'eel order with $\boldsymbol{N}||x$. Two magnon branches are shown in blue and red.}
\label{fig:J}
\end{figure}

 We start with an analysis of the spin model for the spin $S=5/2$: 
\noindent
\begin{displaymath}
\mathcal{H}^{S} = - \sum_{\langle ij \rangle} \left( J_{ij} \boldsymbol{S}_{i} \cdot \boldsymbol{S}_{j} - \boldsymbol{D}_{ij} \cdot [\boldsymbol{S}_{i} \times \boldsymbol{S}_{j}] \right).
\end{displaymath}
\noindent To evaluate the isotropic exchange ($J_{ij}$) and DM ($\boldsymbol{D}_{ij}$) interactions, we first construct a realistic electronic model for the magnetic Mn $3d$ bands based on the electronic structure obtained in LDA (see Fig.~\ref{fig:str}) and using for these purposes Wannier functions technique~\cite{review2008,WannierRevModPhys}. The intraatomic Coulomb repulsion and exchange interaction are evaluated within constrained random-phase approximation as $U=3.6$ eV and $J_{\rm H}=0.9$ eV~\cite{review2008,cRPA}. Then, we solve this model within the mean-field Hartree–Fock (HF) approximation and evaluate the interatomic exchange parameters using linear-response theory, formulated as a perturbation expansion in infinitesimal spin rotations around the AFM ground state~\cite{review2024}. In practice, the model contains eight sizable parameters specifying isotropic interactions, which are illustrated in Fig.~\ref{fig:J}(a). 
\noindent
\noindent Their numerical values are  $J_{1}=-1.0101$, $J_{2x}=-0.0035$, $J_{2z}=-0.8531$, $J_{3}=-0.0137$, $J_{4}=-0.0198$, $\delta J_{4}=-0.0179$, $J_{5}=-0.0033$, and $J_{6}=-0.0119$ meV. The DM vectors $\boldsymbol{D}_{ij}=(\pm d_{x},\pm d_{y},0)$ are shown in Fig.~\ref{fig:J}(b) and their numerical values are $d_{x}=d_{y}=0.0097$ meV. For the $d^5$ configuration of Mn$^{2+}$, the exchange parameters can also be obtained using the superexchange relation $2J_{ij}S^2B = -\sum_{ab} t_{ij}^{ab} t_{ji}^{ba}$ in the strong-coupling regime, with $2B=U+4J_{\rm H}$ ($= 7.2$ eV) the intraatomic spin splitting and $\hat{t}_{ij} = [t_{ij}^{ab}]$ the $5 \times 5$ transfer-integral matrix. The resulting values are very similar to those from the HF calculation; in particular $J_{1}=-0.9967$, $J_{2z}=-0.8360$, $J_{4}=-0.01939$, and $\delta J_{4}=-0.0165$ meV. The important point is that the strong-coupling regime does not differentiate between the various magnetic interactions, in that all exchange parameters scale inversely with $B$. 

\par The interaction $\pm \delta J_{4}$ alternates between the two sublattices and is responsible for the magnon splitting. In general, two main mechanisms contribute to $\pm \delta J_{4}$~\cite{Naka,Naka_Spintronics,JPSJ2009}. The first arises from direct hoppings through conventional superexchange processes, as discussed above. The second originates from super-superexchange paths involving intermediate Mn sites. This contribution appears only in higher orders of $t/B$ (with $t$ the effective hopping between an atom of the bond and the intermediate Mn site) and therefore becomes negligible in the strong-coupling regime realized in MnF$_2$. This is consistent with the above analysis, where the more rigorous linear-response evaluation of the exchange interactions is well reproduced by the superexchange theory. The situation is fundamentally different from the perovskite manganese oxides considered in Ref.~\cite{JPSJ2009}, where the on-site Coulomb repulsion is relatively small due to more efficient screening, thereby activating the super-superexchange processes.

\par Defining the uniaxial anisotropy energy as $K_{u}(S_{z})^2$, we estimate the constant $K_{u}$ to be $-0.0003$ meV, which favors spin alignment along the $z$ axis, in agreement with the experiment. However, a magnetic field of order $H_{\rm SF} \approx \sqrt{2H_{\rm E}H_{\rm A}} \sim 2$ T (where $\mu_{\rm B}H_{\rm E} \approx 4|J_{1}|S$ and $\mu_{\rm B} H_{\rm A}=|K_{u}|S$) should induce a spin-flop transition and rotate the spins into the $xy$ plane. The experimental value $K_{u}=-0.0267$ meV is substantially larger~\cite{MoranoMnF2}, which is somewhat counterintuitive for the nearly spherical $d^{5}$ configuration of Mn$^{2+}$. The origin of such a large $K_{u}$ is generally attributed to intersite dipole-dipole interactions~\cite{MoranoMnF2}. The experimental spin-flop field $H_{\rm SF} \sim 9$ T is also larger, although the discrepancy is only moderate~\cite{FelcherKleb}.

\par Notably, $\delta J_{4}$ is much smaller than $J_{1}$ and $J_{2z}$, and is comparable to other parameters, including the DM interactions. Therefore, it is not expected to play an important role in the magnon dispersion, which can be evaluated following Ref.~\cite{Hoyer}, taking into account the additional exchange interactions shown in Fig.~\ref{fig:J}(a). The resulting magnon dispersion is summarized in Fig.~\ref{fig:J}(c) for the N\'eel vector $\boldsymbol{N} = \frac{\boldsymbol{S}_{2}-\boldsymbol{S}_{1}}{2S}$ oriented along $x$. The splitting of the magnon branches caused by $\delta J_{4}$ is indeed small. 

\par There are several discrepancies relative to the experimental data, which are typically interpreted in terms of two ferromagnetic (FM) interactions, $J_{2x}=0.0044$ and $J_{2z}=0.0677$ meV, and one antiferromagnetic (AFM) one, $J_{1}=-0.3022$ meV~\cite{MoranoMnF2}. Moreover, $T_{\rm N} = 126$ K, evaluated in the random-phase approximation with the obtained parameters $J_{ij}$, exceeds the experimental value by a factor of two. The missing FM contributions to $J_{ij}$ may be related to the magnetic polarization of the F $2p$ states~\cite{review2024}. However, LDA calculations reveal a large gap between the F $2p$ and Mn $3d$ bands (see Fig.~\ref{fig:str}(b)), justifying a separate treatment of the Mn $3d$ manifold as the starting point for analyzing the magnetic properties of MnF$_2$. Furthermore, since F connects antiferromagnetically coupled Mn atoms, the magnetic polarization of F is expected to be weak due to the strong cancellation of contributions with the majority- and minority spin states. On the other hand, since the F ions occupy off-centrosymmetric positions, this cancellation is not complete. We leave this problem for future investigation and conclude this section by noting that the splitting of the chiral magnon modes in Fig.~\ref{fig:J}(c) is small and comparable to experimental estimates, which suggest $\Delta E_{\boldsymbol{k}} < 120$ meV~\cite{MoranoMnF2,FaureMnF2}. 

\section{\label{sec:effel} Effective electronic model}
\par To discuss the electronic properties, one can use the general model including five $3d$ orbitals on each Mn site, as employed in the analysis of the magnetic interactions. Nevertheless, for the half-filled $d^{5}$ configuration of Mn$^{2+}$, it can be more instructive to consider an effective one-orbital model with the parameters derived from the superexchange expressions for $J_{ij}$ and $\boldsymbol{D}_{ij}$:
\noindent
\begin{displaymath}
J_{ij} = -\frac{t_{ij}^2}{2BS^2} \quad \text{and}  \quad \boldsymbol{D}_{ij} = \frac{t_{ij}\boldsymbol{\lambda}_{ij}}{BS^2},
\end{displaymath}
\noindent where $t_{ij}$ is the effective transfer integral and $\boldsymbol{\lambda}_{ij}$ specifies the SO interaction on the bond through $\hat{\mathcal{H}}^{\rm so}_{ij} = i\boldsymbol{\lambda}_{ij}\cdot \hat{\boldsymbol{\sigma}}$ with the vector of Pauli matrices $\hat{\boldsymbol{\sigma}}=(\hat{\sigma}_{x},\hat{\sigma}_{y},\hat{\sigma}_{z})$. Since the exchange parameters obtained within the five-orbital model are well reproduced by the superexchange theory, such a mapping onto the one-orbital model does not introduce any ambiguity. Then, using the parameters $J_{ij}$ and $\boldsymbol{D}_{ij}$ derived above, one readily obtains (up to overall signs that do not affect the optical conductivity): $t_{1}=213.38$, $t_{2x}=12.51$, $t_{2z}=196.10$, $t_{3}=24.86$, $t_{4}=25.28$, $\delta t_{4}=15.95$, $t_{5}=12.16$, $t_{6}=23.18$, and $\lambda_{x}=\lambda_{y}=1.02$ meV. As expected, $\delta t_{4}$ is not particularly strong.

\par After the Fourier transform, the Hamiltonian takes the form:
\noindent
\begin{equation}
\hat{\mathcal{H}}_{\boldsymbol{k}} = h^{\phantom{0}}_{\boldsymbol{k}} + h^{1}_{\boldsymbol{k}} \hat{\tau}_{x} + \boldsymbol{h}^{\rm so}_{\boldsymbol{k}} \cdot \hat{\boldsymbol{\sigma}} \hat{\tau}_{y}  + (\delta h^{\phantom{0}}_{\boldsymbol{k}} + B \boldsymbol{N} \cdot \hat{\boldsymbol{\sigma}} ) \hat{\tau}_{z} ,
\label{eq:Hk}
\end{equation}
\noindent where $\hat{\boldsymbol{\tau}} = (\hat{\tau}_{x},\hat{\tau}_{y},\hat{\tau}_{z})$ are pseudospin Pauli matrices describing two sublattices~\cite{Roig}, $h^{\phantom{0}}_{\boldsymbol{k}}=2t_{2x}(\cos k_{x}+\cos k_{y}) + 2t_{2z}\cos k_{z} + 4t_{3}(\cos k_{x}+\cos k_{y})\cos k_{z} + 4 t_{4} \cos k_{x} \cos k_{y} + 2t_{6}\cos2k_{z}$, $\delta h^{\phantom{0}}_{\boldsymbol{k}} = 4\delta t_{4} \sin k_{x} \sin k_{y}$, $h^{1}_{\boldsymbol{k}} = 8t_{1}\cos\frac{k_{x}}{2}\cos\frac{k_{y}}{2}\cos\frac{k_{z}}{2} + 8t_{5}\cos\frac{k_{x}}{2}\cos\frac{k_{y}}{2}\cos\frac{3k_{z}}{2}$, $\boldsymbol{h}^{\rm so}_{\boldsymbol{k}}=(h^{x}_{\boldsymbol{k}},h^{y}_{\boldsymbol{k}},0)$, $h^{x}_{\boldsymbol{k}}=8\lambda_{x} \sin\frac{k_{x}}{2} \cos\frac{k_{2}}{2} \sin\frac{k_{z}}{2}$, 
$h^{y}_{\boldsymbol{k}}=-8\lambda_{y} \cos\frac{k_{x}}{2} \sin\frac{k_{2}}{2} \sin\frac{k_{z}}{2}$, and $\boldsymbol{N} = (\cos \varphi, \sin \varphi,0)$ specifies the direction of the N\'eel vector in the $xy$ plane. Because the atomic sites coincide with inversion centers, the inversion operation enforces $\hat{\mathcal{H}}_{-\boldsymbol{k}}=\hat{\mathcal{H}}_{\boldsymbol{k}}$.

\par The diagonalization of $\hat{\mathcal{H}}_{\boldsymbol{k}}$ yields the following eigenvalues~\cite{Roig}: 
\noindent
\begin{displaymath}
\varepsilon^{\sigma}_{\boldsymbol{k},\nu} = h^{\phantom{0}}_{\boldsymbol{k}} + \nu \sqrt{ B^{2}+(\delta h^{\phantom{4}}_{\boldsymbol{k}})^{2} + ( h^{1}_{\boldsymbol{k}})^{2} + (\boldsymbol{h}^{\rm so}_{\boldsymbol{k}})^{2} +2\sigma B \sqrt{(\delta h^{\phantom{4}}_{\boldsymbol{k}})^{2} + |\boldsymbol{h}^{\rm so}_{\boldsymbol{k}} \times \boldsymbol{N}|^{2}} }, 
\end{displaymath}
\noindent where $\nu = \pm$ and $\sigma = \pm$ are band and spin indices, respectively. Without SO interaction, the condition $\delta h^{\phantom{4}}_{\boldsymbol{k}}=0$ defines altermagnetic nodal surfaces of spin-degenerate bands without SO coupling~\cite{Roig,SmejkalSA}. If $\boldsymbol{h}^{\rm so}_{\boldsymbol{k}}$ is not parallel to $\boldsymbol{N}$, the SO interaction lifts the spin degeneracy. However, the term $\boldsymbol{h}^{\rm so}_{\boldsymbol{k}} \times \boldsymbol{N}$ does not contribute to AHE or magneto-optical response and, therefore, can be neglected. In particular, only the components of $\boldsymbol{h}^{\rm so}_{\boldsymbol{k}}$ parallel to $\boldsymbol{N}$ contribute to AHE~\cite{HiddenFM,NakaOrganic,arXiv2025}. Therefore, although the $P4_{2}/mnm$ symmetry enforces $\lambda_{x}=\lambda_{y}$, for practical purposes one may retain only the component of $\boldsymbol{\lambda}$ parallel to $\boldsymbol{N}$ and neglect the other. For instance, if $\boldsymbol{N}||x$, one can set $\lambda_{y}=0$. In this case, the states with $\sigma = \pm$ become decoupled and for large $B$ we will have the following expression for $\varepsilon^{\sigma}_{\boldsymbol{k},\nu}$:
\noindent
\begin{equation}
\varepsilon^{\sigma}_{\boldsymbol{k},\nu} \approx h^{\phantom{0}}_{\boldsymbol{k}} + \nu \left(B + \sigma \delta h^{\phantom{0}}_{\boldsymbol{k}} +\frac{(h^{1}_{\boldsymbol{k}})^{2}}{2B} \right).
\label{eq:eklargeB}
\end{equation}
\noindent When $\delta h^{\phantom{4}}_{\boldsymbol{k}}=0$, the bands are spin-degenerate.

\par Since $\boldsymbol{N}$ and $\boldsymbol{h}^{\rm so}_{\boldsymbol{k}}$ in Eq.~(\ref{eq:Hk}) lie in the $xy$ plane, the time-reversal operation can be written as $\mathcal{T}=\mathcal{S}K$, where $\mathcal{S}=i\hat{\sigma}_{z}$ represents a 180$^{\circ}$ spin rotation about $z$ and $K$ denotes complex conjugation. Furthermore, the lattice translation by $\mathbf{t}$ in Eq.~(\ref{eq:Hk}) is equivalent to a 180$^{\circ}$ rotation of $\hat{\boldsymbol{\tau}}$ about $x$, which transforms $\hat{\tau}_{y} \to -\hat{\tau}_{y}$, $\hat{\tau}_{z} \to -\hat{\tau}_{z}$, while leaving $\hat{\tau}_{x}$ invariant. Then, Eq.~(\ref{eq:Hk}) exhibits two distinct sources of time-reversal symmetry breaking: 
\noindent
\begin{enumerate}
\item $\boldsymbol{h}^{\rm so}_{\boldsymbol{k}}$ is finite but $\delta h^{\phantom{0}}_{\boldsymbol{k}}=0$. In this case, the Hamiltonian (\ref{eq:Hk}) is invariant under $\{ \mathcal{S} | \mathbf{t} \}$, while $\mathcal{T}$ and $\{ \mathcal{T} | \mathbf{t} \}$ symmetries are broken~\cite{arXiv2025}. The $\{ \mathcal{S} | \mathbf{t} \}$ symmetry enables the use of the generalized Bloch theorem~\cite{Sandratskii_review}, showing that this unconventional antiferromagnet is equivalent to a ferromagnet on a reduced unit cell, specified by the translations $\mathbf{t}_{1}=(-\frac{1}{2},\frac{1}{2},\frac{1}{2})$, $\mathbf{t}_{2}=(\frac{1}{2},-\frac{1}{2},\frac{1}{2})$, and $\mathbf{t}_{3}=(\frac{1}{2},\frac{1}{2},-\frac{1}{2})$, with one magnetic site~\cite{HiddenFM}. This naturally explains the appearance of AHE and other ``ferromagnetic'' phenomena, which are odd in the SO coupling. The property holds even for $\boldsymbol{h}^{\rm so}_{\boldsymbol{k}} \times \boldsymbol{N} \ne 0$, when the band degeneracy is lifted by the SO interaction. Nevertheless, the bands can still be unfolded into the larger Brillouin zone~\cite{HiddenFM}. 
\item $\delta h^{\phantom{0}}_{\boldsymbol{k}}$ is finite but $\boldsymbol{h}^{\rm so}_{\boldsymbol{k}}=0$. Then, the Hamiltonian (\ref{eq:Hk}) is real and $K$-invariant, although $\mathcal{T}$ (and $\mathcal{T}$ combined with $\mathbf{t}$) remains broken. Besides regular AFM order parameter, such state can be described in terms of ferroically ordered spin magnetic octupoles~\cite{McClartyRau,BhowalSpaldin,SatoHayami,OikePetersShinada}. Nevertheless, AHE and net orbital magnetization will vanish.   
\end{enumerate}
    
\par Thus, band splitting alone cannot induce the AHE or magneto-optical effects, but it can modulate their magnitude when combined with the SO coupling.

\subsection{\label{sec:cond} Conductivity tensor and properties}
\par The elements of conductivity tensor can be evaluated using the Kubo formula~\cite{NakaOrganic,Oppeneer}:
\noindent
\begin{equation}
\sigma_{\alpha \beta}(\omega) = -i \hbar \sum_{lm} \int_{\mathrm{BZ}} \frac{d \boldsymbol{k}}{(2 \pi)^3} \frac{f(\varepsilon_{\boldsymbol{k},m})-f(\varepsilon_{\boldsymbol{k},l})}{\varepsilon_{\boldsymbol{k},m}-\varepsilon_{\boldsymbol{k},l}} \frac{J_{\boldsymbol{k},ml}^{\alpha}J_{\boldsymbol{k},lm}^{\beta}}{\hbar \omega + \varepsilon_{\boldsymbol{k},m}-\varepsilon_{\boldsymbol{k},l} + i\delta },
\label{eq:cond}
\end{equation}
\noindent where $l$ and $m$ are composite $(\nu,\sigma)$ indices, $f(\varepsilon_{\boldsymbol{k},l})$ is the Fermi-Dirac distribution function, the integration goes over the 1st Brillouin zone (BZ),  $J_{\boldsymbol{k},ml}^{\alpha}$ is the matrix element of electric current operator, $\hat{J}^{\alpha}_{\boldsymbol{k}} = -\frac{e}{\hbar} \frac{\partial \hat{\mathcal{H}}_{\boldsymbol{k}} }{\partial k_{\alpha}}$, between states $l$ and $m$, and $\delta$ is a positive infinitesimal. Because the states $l$ and $m$ are orthogonal, $h^{\phantom{0}}_{\boldsymbol{k}}$ does not contribute to $J_{\boldsymbol{k},ml}^{\alpha}$. However, it does affect $f(\varepsilon_{\boldsymbol{k},l})$, and therefore influences the shape of the Fermi surface~\cite{NakaOrganic,arXiv2025}.

\subsubsection{Anomalous Hall Effect}
\par
\begin{figure}[tbh]
\centering
\includegraphics[width=15.0cm]{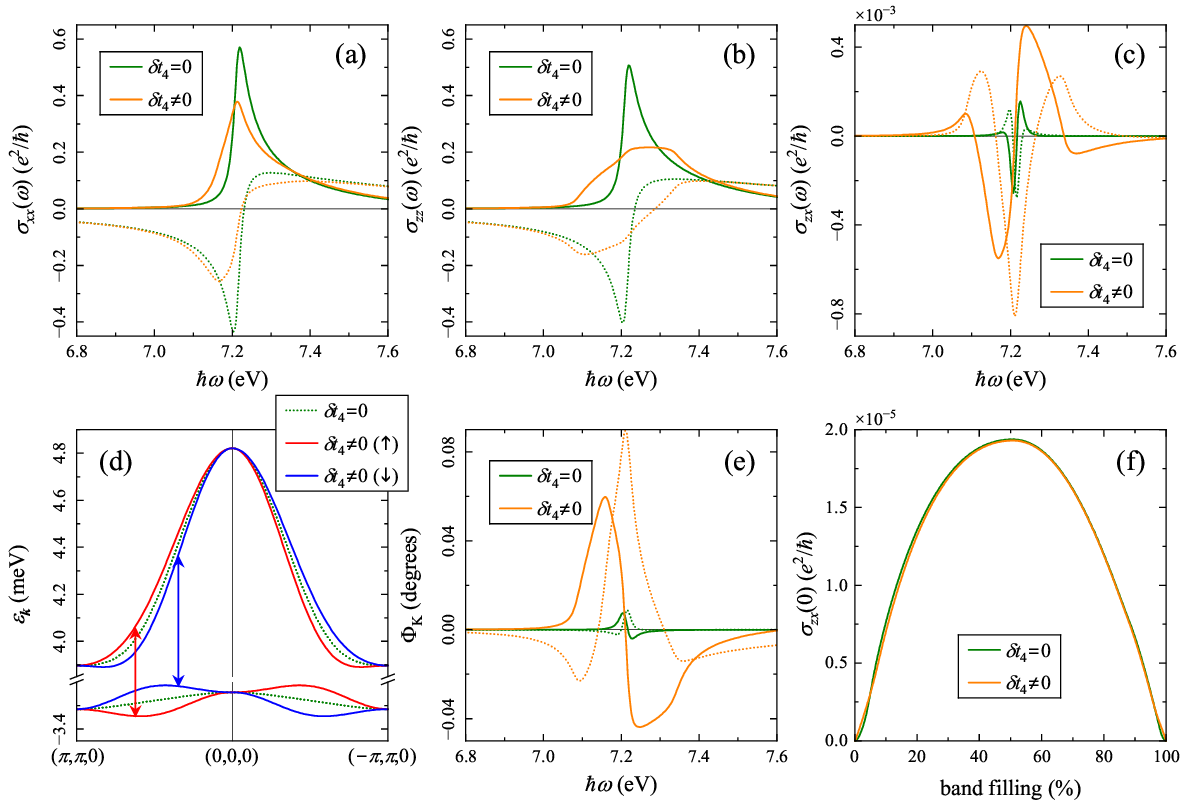}
\caption{Summary of optical, magneto-optical, and Hall responses for $\boldsymbol{N}||x$ with and without the altermagnetic band splitting ($\delta t_{4}$). (a), (b), and (c) Real (solid line) and imaginary (dotted line) parts of $\sigma_{xx}(\omega)$, $\sigma_{zz}(\omega)$, and $\sigma_{zx}(\omega)=-\sigma_{xz}(\omega)$. (d) Example of band dispersion. The allowed optical transitions are shown by arrows. (e) Complex Kerr effect $\Phi_{\mathrm{K}} = \phi_{\mathrm{K}} + i \epsilon_{\mathrm{K}}$. Kerr rotation angle ($\phi_{\rm K}$) and ellipticity ($\epsilon_{\rm K}$) are shown by the solid and dotted lines, respectively. (f) AHE versus the population of the low-energy band.}
\label{fig:mo}
\end{figure}

 The AHE is determined by $\sigma_{\alpha \beta}(0)$, which can be recast in the form~\cite{NagaosaRevModPhys}: 
\begin{displaymath}
\sigma_{\alpha \beta}(0) = - \frac{e^2}{\hbar} \sum_{l} \int_{\mathrm{BZ}} \frac{d \boldsymbol{k}}{(2 \pi)^3} f(\varepsilon_{\boldsymbol{k},l}) \, \Omega_{\boldsymbol{k},l}^{\gamma} ,
\end{displaymath}
\noindent where $\Omega_{\boldsymbol{k},l}^{\gamma} = i \langle \partial_{k_{\alpha}} u_{\boldsymbol{k},l} | \partial_{k_{\beta}} u_{\boldsymbol{k},l} \rangle $ is the Berry curvature, $u_{\boldsymbol{k},l}$ is the eigenstate of  $\hat{\mathcal{H}}_{\boldsymbol{k}}$, and $(\alpha, \beta, \gamma)$ is an even permutation of $(x, y, z)$. In this case, all $\delta t_{4}$-dependence of $\Omega_{\boldsymbol{k},l}^{\gamma}$ is solely given by $u_{\boldsymbol{k},l}$. For large $B$, one can identify two main contributions to $\Omega_{\boldsymbol{k},l}^{\gamma}$~\cite{arXiv2025}. The first one ($\Omega_{\boldsymbol{k},l}^{\gamma, o}$, which is odd in $B$) is proportional to $\frac{t_{1} \lambda_{x}}{B|B|}$ (for $\boldsymbol{N}||x$) and does not depend on $\delta t_{4}$. The second contribution ($\Omega_{\boldsymbol{k},l}^{\gamma, e}$) is even in $B$ and proportional to $\frac{t_{1} \lambda_{x} \delta t_{4}}{|B|^3}$. Since $\frac{\Omega_{\boldsymbol{k},l}^{\gamma, e}}{\Omega_{\boldsymbol{k},l}^{\gamma, o}} \sim \frac{\delta t_{4}}{B} \sim 0.002$, the band splitting provides only a small correction to $\Omega_{\boldsymbol{k},l}^{\gamma, o}$, existing in the spin-degenerate case. The typical example is shown in Fig.~\ref{fig:mo}(f), provided that the metallic state in MnF$_2$ can be realized via the hole doping and the corresponding doping-dependence can be described in the rigid-band approximation. 
\noindent
\noindent One can clearly see that the contribution of $\delta t_{4}$ to AHE is negligible.

\subsubsection{Optical and magneto-optical response}
\par For large $B$, Eq.~(\ref{eq:cond}) reduces to 
\noindent
\begin{equation}
\sigma_{\alpha \beta}(\omega) \approx \frac{i \hbar}{2B} \sum_{\sigma} \int_{\mathrm{BZ}} \frac{d \boldsymbol{k}}{(2 \pi)^3} \frac{\textrm{Re} [J_{\boldsymbol{k},lm}^{\alpha}J_{\boldsymbol{k},ml}^{\beta}] + i \textrm{Im} [J_{\boldsymbol{k},lm}^{\alpha}J_{\boldsymbol{k},ml}^{\beta}]}{\hbar \Delta \omega -2\sigma \delta h^{\phantom{0}}_{\boldsymbol{k}} - \dfrac{(h^{1}_{\boldsymbol{k}})^{2}}{B} + i\delta },
\label{eq:condlargeB}
\end{equation}
\noindent where $\hbar \Delta \omega = \hbar \omega - 2B$, $l$ and $m$ are the states with $(-,\sigma)$ and $(+,\sigma)$, respectively, and we used Eq.~(\ref{eq:eklargeB}) to express $\varepsilon_{\boldsymbol{k},m}-\varepsilon_{\boldsymbol{k},l}$. The real part of $J_{\boldsymbol{k},lm}^{\alpha}J_{\boldsymbol{k},ml}^{\beta}$ in Eq.~(\ref{eq:condlargeB}) contributes to the diagonal components of the conductivity tensor, whereas the imaginary part determines the off-diagonal conductivity, $\sigma_{\alpha \beta}(\omega) = - \sigma_{\beta \alpha}(\omega)$. All transitions occur between states with the same $\sigma$ as explained in Fig.~\ref{fig:mo}(d).

\par Although $\delta t_{4} \sim 16$ meV is much smaller than $t_{1} \sim 213$ meV, it contributes to the energy difference directly, whereas the contribution from $t_{1}$ appears only at order $\frac{(t_{1})^2}{B} \sim 6$ meV. The essential distinction is that $t_{1}$ acts between antiferromagnetically coupled sublattices, while $\delta t_{4}$ acts within the same sublattice and therefore its contribution to the eigenvalues in Eq.~(\ref{eq:eklargeB}) is not suppressed by large $B$. Thus, unlike in the case of AHE, $\delta t_{4}$ will largely control the optical properties of MnF$_2$ at energies $\hbar \omega \sim 2B$. This behavior also contrasts with the magnon spectrum, where all contributions scale as $1/B$.  This effect is not related to the SO interaction and can be clearly seen in both the diagonal and off-diagonal components of the optical conductivity in Fig.~\ref{fig:mo}(a)-(c). First, the bandwidth of the optical excitations is of the order of $16 \delta t_{4} \sim 0.3$ eV and is primarily determined by $\delta t_{4}$. Second, the shape of these excitations also depends on $\delta t_{4}$. For realistic values of $\delta t_{4} \sim 16$ meV, obtained from electronic structure calculations, the spectral weight is centered \emph{around} $\hbar \omega_{0} = 2B$. In contrast, when $\delta t_{4} = 0$, the shape of $\sigma_{\alpha \beta}(\omega)$ is controlled by $\frac{(h^{1}_{\boldsymbol{k}})^{2}}{B}$, which is \emph{positive}, so all spectral weight shifts \emph{above} $\hbar \omega_{0}$. Third, the BZ integral depends strongly on the energy dispersion appearing in the denominator of Eq.~\ref{eq:condlargeB}. The diagonal conductivity satisfies the f-sum rule, which means that $t_{1}$ and $\delta t_{4}$ can reshape $\mathrm{Re}[\sigma_{\alpha \alpha}(\omega)]$ but cannot change its total spectral weight, as is clearly seen in Fig.~\ref{fig:mo}(a),(b). By contrast, the off-diagonal conductivity is more flexible and its magnitude is strongly enhanced by $\delta t_{4}$.

\par Finally, we evaluate the complex Kerr effect using the formula 
\noindent
\begin{displaymath}
\Phi_{\mathrm{K}} = \phi_{\mathrm{K}} + i \epsilon_{\mathrm{K}} = \frac{-\sigma_{zx}(\omega)}{\sigma_{zz}(\omega)\sqrt{1 + \dfrac{4\pi i}{\omega}\sigma_{zz}(\omega)}}, 
\end{displaymath}
\noindent which becomes exact for cubic materials with net magnetization along $y$~\cite{Oppeneer}, corresponding to $\boldsymbol{N}||x$. The results are shown in Fig.~\ref{fig:mo}(e). $\Phi_{\mathrm{K}}$ is finite without $\delta t_{4}$, showing that it originates from relativistic SO coupling, as in the AHE. However, unlike the AHE, $\Phi_{\mathrm{K}}$ is highly sensitive to $\delta t_{4}$, which strongly enhances its magnitude. The expected Kerr rotation angle $\phi_{\rm K} = \mathrm{Re} [\Phi_{\mathrm{K}}]$ in AFM MnF$_2$ is still about one order of magnitude smaller than in typical ferromagnets, such as bcc Fe and fcc Ni~\cite{Oppeneer}. Nevertheless, this example clearly shows that even a relatively small band splitting, $\delta t_{4}$, can strongly modify $\Phi_{\mathrm{K}}$ and thereby provide a means to control the magneto-optical properties of MnF$_2$ and other altermagnetic materials.   

\section{\label{sec:summary} Summary and Conclusions}
\par Recently, altermagnetism has attracted considerable attention owing to the possibility of realizing spin-split electronic bands in nearly collinear antiferromagnets. Much of the current research effort is directed toward identifying materials with large altermagnetic band splitting, reflecting the widespread expectation that such splitting plays a central role in the time-reversal-symmetry-breaking responses of these systems. When the experimentally inferred splitting turns out to be small, the result is often viewed as unexpected, and the material is typically regarded as less promising.

\par In this context, it is instructive to revisit Turov’s original concept of centrosymmetric antiferromagnetism, formulated several decades ago to describe the properties of AFM materials on antipolar lattices~\cite{TurovBook}. This concept is very general and is not restricted to the altermagnetic splitting of electronic bands. Although Turov did not explicitly discuss such microscopic band-structure effects -- focusing instead on macroscopic manifestations of time-reversal-symmetry breaking -- their existence follows naturally from symmetry considerations~\cite{Sandratskii1981,SmejkalPRX1}. At the same time, symmetry alone does not determine whether the resulting band splitting is large or small. It may be accidentally small, or even vanish by symmetry, as in the one-orbital model of La$_2$CuO$_4$, which is widely used in the theory of superconductivity~\cite{arXiv2025}. Nevertheless, beyond the altermagnetic splitting of bands, there exist other important microscopic features inherent to centrosymmetric antiferromagnets that have been largely overshadowed by the current focus on altermagnetism. In particular, spatial inversion in antipolar materials imposes a symmetry constraint on the SO interaction, causing it to transform as an AFM object and to collaborate with the N\'eel field~\cite{arXiv2025,HiddenFM}. Because both quantities obey the same $\{ \mathcal{S} | \mathbf{t} \}$ symmetry, the system can, in the local frame, be viewed as an effective ferromagnet with one magnetic sublattice~\cite{HiddenFM}. This naturally explains why centrosymmetric antiferromagnets can exhibit AHE, magneto-optical responses, and other phenomena that are odd in the SO interaction. All of these effects can exist even in the absence of altermagnetic band splitting. However, they all disappear without spin-orbit coupling, which is indispensable for their emergence.

\par Thus, altermagnetic band splitting does not play a primary role in the emergence of SO-odd properties. However, it can substantially influence their magnitude. This raises a natural question: what do we actually mean when we say that the altermagnetic splitting is ``large'' or ``small''? The answer depends on the specific physical property under consideration, rather than on the material itself. For instance, in MnF$_2$ the effect of altermagnetic splitting on the magnon spectra and on the anomalous Hall effect (under doping) is indeed negligible. However, it can significantly reshape the optical conductivity and strongly boost the magneto-optical response. A ``large'' or ``small'' altermagnetic splitting observed in one type of measurement therefore says little about how promising a material is overall. Everything depends on the property: for some responses the effect may be minor, while for others it can be substantial.

\section*{Acknowledgement}
\par MANA is supported by World Premier International Research Center Initiative (WPI), MEXT, Japan.

\end{document}